# ASCA observations of high redshift quasars

J. Siebert[1,2], M. Matsuoka[2], W. Brinkmann[1,2], M. Cappi[2], T. Mihara[2], and T. Takahashi[3]

[1] Max–Planck–Institut für Extraterrestrische Physik, Giessenbachstrasse, D-85740 Garching, FRG
[2] Institute of Physical and Chemical Research (RIKEN), Hirosawa 2-1, Wako, Saitama 351-01, Japan
[3] Institute of Space and Astronautical Science, Yoshinodai 3-1-1, Sagamihara, Kanagawa 229, Japan

**Abstract.** ASCA observations of 4 high redshift radio–loud quasars with $1.44<z<3.21$ are presented. The spectral analysis for three of them (PKS 0332-403, PKS 0537-286, PKS 2149-306) reveals that their X-ray continuum emission is well represented by a simple power–law model plus absorption with photon indices of $\Gamma = 1.92^{+0.30}_{-0.20}$ (PKS 0332-403), $\Gamma = 1.63^{+0.14}_{-0.12}$ (PKS 0537-286) and $\Gamma = 1.57 \pm 0.05$ (PKS 2149-306). The fourth and most distant object, PKS 1614+051 at z=3.21, was detected, but a detailed spectral analysis is impossible due to the small number of photons. We find evidence for excess absorption above the Galactic $N_H$–value in the ASCA data of PKS 2149-306, which is not confirmed by the ROSAT All-Sky Survey PSPC spectrum of this source. This could probably be due to variable absorption. The ROSAT spectrum of PKS 0537-286, deduced from a 10 ksec pointed PSPC observation, is consistent with the ASCA results. Thermal bremsstrahlung models also give acceptable fits to the ASCA data with best fit (rest frame) temperatures of 10.4, 33.5 and 45.8 keV for PKS 0332-403, PKS 0537-286 and PKS 2149-306, respectively. More complicated models for the X-ray continuum are not required, in particular, tight upper limits on the strength of the Fe-K emission line are given. The broad band spectral energy distributions from the radio to the $\gamma$-rays are presented and discussed.

## 1. Introduction

The study of X-ray spectra with reasonable signal-to-noise ratios for a large number of high redshift quasars has become feasible with recent X-ray satellites, which combine large collecting areas with sufficient angular and spectral resolution (ROSAT, Trümper 1984; ASCA, Tanaka et al. 1994). The *Gas Imaging Spectrometer* (GIS, Ohashi et al. 1991) and the *Solid State Imaging Spectrometer* (SIS, Burke et al. 1991) on board ASCA are particularly well suited for this purpose as they provide excellent spectral resolution in a broad energy band (0.4–10 keV). By combining ASCA data with information from ROSAT PSPC observations, we are able to study X-ray spectra from 0.4 keV up to 40 keV in the rest frame of high redshift quasars ($z\sim3$).

These unprecedented capabilities allow us to address important questions such as the evolution of the quasar X-ray spectra with time and/or luminosity (Canizares & White 1989, Della Ceca & Maccacaro 1991, Elvis et al. 1994a). Current studies of the X-ray luminosity function of quasars (e.g. Boyle et al. 1993) indicate that these objects are typically 50 to 100 times brighter in X-rays at $z\sim3$ compared to z=0. Thus, the study of the X-ray spectra of high-redshift quasars can also contribute to our understanding of the X-ray emission processes involved in these objects.

Further, the presence or absence of absorption at soft X-ray energies in excess of the Galactic value in high redshift quasars enables us to draw conclusions on the properties of the suggested intervening absorbers or the intergalactic medium via an X-ray Gunn-Peterson test (Elvis et al. 1994a, Aldcroft et al. 1994). Finally, the spectral signatures of the presently discussed reflection models for AGN (e.g. Lightman & White 1988), in particular the predicted 6.4 keV Fe-K fluorescence line from cold matter, are redshifted to energies at which ASCA has its highest effective area.

In this paper we present the results of the ASCA observations of four high redshift radio-loud quasars, namely PKS 0332-403, PKS 0537-286, PKS 1614+051 and PKS 2149-306. The data are supplemented by a pointed ROSAT PSPC observation of PKS 0537-286, and ROSAT All-Sky Survey observations for the other objects. Table 1 summarizes some basic properties of the objects.

Throughout the paper we assume $H_0 = 50$ km sec$^{-1}$ Mpc$^{-1}$ and $q_0 = 0$.

Details of the ASCA observations are given in Table 2. All objects were observed in 1-CCD faint mode with both SIS detectors, except for PKS 0537-286 for which half of the observation was performed in 2-CCD mode. All sources were at the nominal positions in the SIS and GIS, i.e., $\sim 5'$ offset from the detector center in both instruments. Unfortunately, the GIS3 observation of PKS 0537-286, taken on March 15, 1994, was affected by an on board software related problem, which resulted in a dramatic loss of spectral resolution. Even after applying the recommended repair procedures, the spectral analysis gave results which differed significantly from all other detectors. We therefore decided to exclude these data from our analysis.

**Table 1.** Basic source properties

| Object | z | $m_v$ | $f_{5GHz}$ Jy | $\alpha_{2.7}^{5}$ [†] | $f_\gamma$ [‡] |
|---|---|---|---|---|---|
| PKS 0332−403 | 1.445 | 18.5 | 2.60 | −0.45 | <1.1 |
| PKS 0537−286 | 3.11 | 20.0 | 0.99 | −0.52 | <0.7 |
| PKS 1614+051 | 3.21 | 19.5 | 0.92 | −0.51 | − |
| PKS 2149−306 | 2.345 | 18.4 | 1.15 | +0.22 | <0.8 |

[†] $F_\nu \propto \nu^{-\alpha}$
[‡] EGRET flux > 100 MeV in $10^{-7}$ photons cm$^{-2}$ s$^{-1}$ from Fichtel et al. (1994)

The standard screening criteria were applied to the data. The minimum elevation angle above the Earth's limb was chosen to be $5°$. To avoid atmospheric contamination, data were only accepted in the SIS when the angle between the target and the day-night terminator was greater than $25°$. Further, a minimum cut-off rigidity of 8 GeV/c and 7 GeV/c was applied for the SIS and the GIS, respectively.

Source counts were extracted from a circular region centered on the targets with a radius of $6'$ for the GIS and $4'$ for the SIS. In view of the difficulties in extracting a local background in 1-CCD observations, we decided to estimate the background for the SIS from the blank-sky event files only.

The choice of the background can influence the spectral results for weak sources. Therefore we compared various methods of background extraction for the GIS. Of course, a local background would be the most reasonable choice, because it is obtained with the same selection criteria as the source. However, as the local background has almost inevitably to be taken at off-axis angles different from the source, it will always suffer from the energy dependent change of the effective area for increasing off-axis angles. This change in effective area cannot be corrected investigated case to case.

As noted by the ASCA team (GIS News Dec.14, 1994), the GIS blank-sky event files are contaminated by NGC 6552, a Seyfert 2 galaxy with a very strong Fe-K emission line (Fukazawa et al. 1994). Unfortunately, this source is in the same detector region as the sources in all our observations. Therefore, also the standard method for background estimation, consisting of extracting the GIS blank-sky background from the same area as that of the source had to be investigated carefully.

**Table 2.** ASCA observation log

| Object | Date 1994 | T[†] ksec | Source counts | | | |
|---|---|---|---|---|---|---|
| | | | SIS0 | SIS1 | GIS2 | GIS3 |
| 0332−403 | 8/12 | 18 | 510 | 385 | 326 | 403 |
| 0537−286 | 3/15 | 29 | 1194 | 882 | 831 | −[‡] |
| 1614+051 | 8/2 | 40 | 452 | 333 | 280 | 340 |
| 2149−306 | 10/26 | 19 | 4616 | 3831 | 4097 | 4404 |

[†] Total effective exposure averaged over all 4 detectors.
[‡] GIS3 data for PKS 0537-286 were excluded. See text.

The following backgrounds were chosen for comparison: a local background in an annulus centered on the source with inner radius $9'$ and outer radius $12'$, a circular local background offset from the source, but at the same off-axis angle, the blank-sky background at the source, the blank-sky background at the source with NGC 6552 excluded, and the blank-sky background offset from the source, but at the same off-axis angle. In general, all methods give results consistent within the errors. For example, in the case of PKS 2149-306, the spectral fits with the different backgrounds agree within 2%. The differences increase with decreasing source intensity, but so do the errors on the spectral parameters.

For PKS 2149-306 we find, however, that the local background is higher than the blank-sky background. Although there is a second source visible at a distance of $\sim 10'$, the contamination of our data is negligible, as the intensity of the second source is only $\sim 2\%$ of PKS 2149-306 and it is also outside the extraction radii for the source and the various local backgrounds. A spectral analysis for this source is impossible due to the small number of source photons. We also failed to optically identify this source from available databases. We tried to determine the spectrum of the excess local background by fitting it with the blank-sky from the same region as background. As the excess is only small in absolute photon numbers (<100 photons), the photon statistics is obviously bad, but a power-law spectrum with an index of $\sim 1.5$ gives an acceptable fit to the data. The spectrum seems to be that the local background is still contaminated by source photons.

As a conclusion from the analysis above, we decided to use the blank-sky background at the source position with NGC 6552 excluded in the spectral analysis of the GIS data.

All spectra were rebinned to have at least 20 photons in each energy channel. This allows the use of the $\chi^2$ technique to obtain the best fit values of the source spectra.

We also looked for variability in the countrates of the sources in the ASCA observations, but failed to find any significant variations.

## 3. Spectral analysis

PKS 1614+051, the most distant quasar in our sample at z = 3.21, turned out to be too weak to allow a spectral analysis. Therefore, we only report the results of the remaining 3 objects here.

We first fitted simple power-law models, modified by neutral absorption with photoelectric absorption coefficients from Morrison & McCammon (1983). Fits were performed for both fixed Galactic absorption and free absorption. The results are summarized in Table 3.

In general the source spectra are well fitted by a simple power-law model. In Fig.1 we show the data, best fit power-law model and the residuals for PKS 0332-403, PKS 0537-286 and PKS 2149-306. The photon indices turn out to be quite flat for PKS 0537-286 ($\Gamma \simeq 1.63^{+0.14}_{-0.12}$) and PKS 2149-306 ($\Gamma \simeq 1.57 \pm 0.05$), while PKS 0332-403 seems to have slightly steeper X-ray continuum emission with a best fit photon index of $\simeq 1.92^{+0.30}_{-0.20}$.

We find evidence for absorption in excess of the Galactic value in PKS 2149-306, which is also the X-ray brightest source in our sample. The 90% lower limit on the $N_H$-value is $0.40 \times 10^{21}$ cm$^{-2}$, which is well above the Galactic value of $0.245 \times 10^{21}$ cm$^{-2}$. This result will be discussed further in § 4.

The best fit spectra of PKS 0332-403 and PKS 0537-286 are consistent with Galactic absorption only. However, we cannot exclude excess absorption in these two quasars based on the ASCA observations alone, as their Galactic $N_H$-values are well below $10^{21}$ cm$^{-2}$ and the GIS as well as the SIS detectors are not sensitive enough at soft energies to measure such low values of absorption accurately enough in weak sources. In the case of PKS 0537-286, where a pointed ROSAT observation is available (see § 4), no evidence for excess absorption is found.

Fitting the spectra with a thermal bremsstrahlung model also gives acceptable fits to the data. The best fit rest frame temperatures are $kT = 33.5^{+12.0}_{-7.6}$ keV and $kT = 45.8^{+9.8}_{-6.6}$ keV for PKS 0537-286 and PKS 2149-306, respectively. As already noted in Elvis et al. (1994b) and Serlemitsos et al. (1994) this would indicate, assuming that low redshift quasars exhibit the same spectral characteristics, that these objects could contribute significantly to the cosmic X-ray background, which can be described by a thermal plasma at a temperature of 40 keV in this energy range. Note, however, that the best-fit temperature for PKS 0332-403, the quasar with the smallest redshift in the sample, is very low with $kT = 10.4^{+4.3}_{-2.6}$ keV.

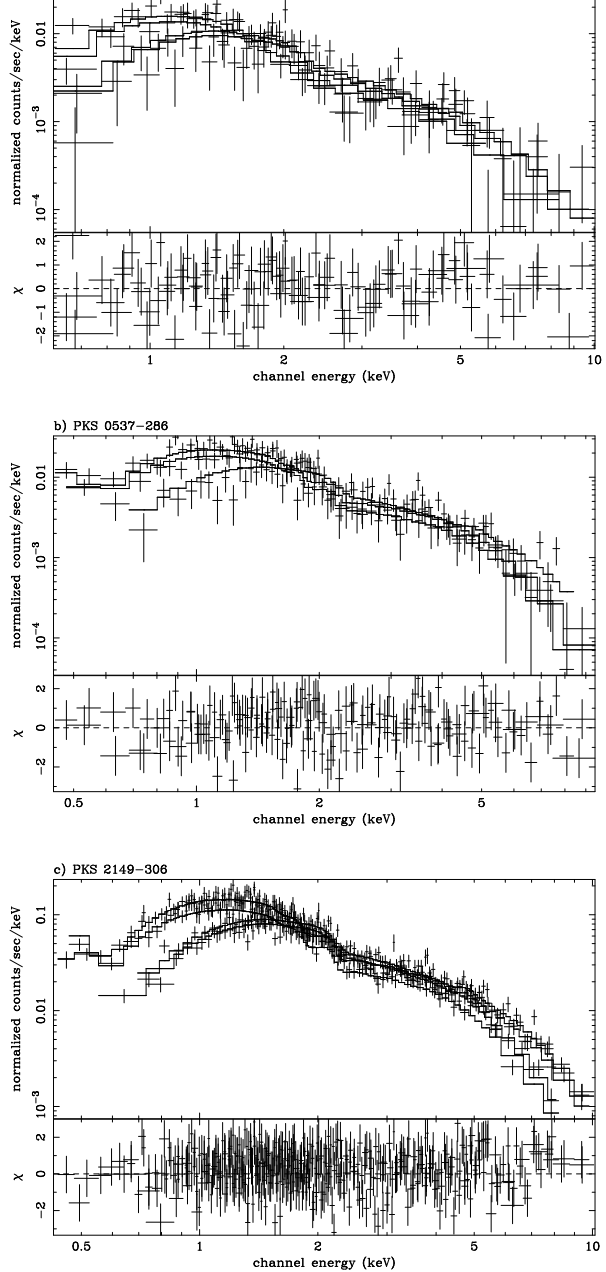

**Fig. 1.** Best fit power-law model and residuals for (a) PKS 0332-403, (b) PKS 0537-286, (c) PKS 2149-306. All detectors were fitted simultaneously.

We further want to draw attention to the fact that temperatures of the order of 40 keV are shifted to 10 keV in the observers frame for redshifts of ~3, which corresponds to the high energy cutoff of ASCA. Given the observed flatness of the ASCA spectra, thermal bremsstrahlung fits will tend to converge at these temperatures. This makes an exact temperature determination rather unreliable.

**Table 3.** Results of the ASCA spectral analysis

| | Absorbed power-law | | | | | |
|---|---|---|---|---|---|---|
| | $\Gamma$ | $N_H^a$ | $\chi^2_{red}$ (dof) | $\Gamma$ | $N_{H,gal}$ | $\chi^2_{red}$ (dof) |
| *PKS 0332-403* | | | | | | |
| SIS0+1 | $2.18^{+0.50}_{-0.36}$ | $< 2.62$ | 1.26(64) | $1.98^{+0.15}_{-0.13}$ | 0.145 | 1.27(65) |
| GIS2+3 | $2.01^{+0.57}_{-0.43}$ | $3.73^{+5.73}_{-3.42}$ | 0.94(68) | $1.64^{+0.18}_{-0.16}$ | 0.145 | 1.00(68) |
| GIS+SIS | $1.92^{+0.30}_{-0.20}$ | $< 1.61$ | 1.20(134) | $1.88^{+0.11}_{-0.10}$ | 0.145 | 1.19(135) |
| *PKS 0537-286* | | | | | | |
| SIS0+1 | $1.64^{+0.18}_{-0.16}$ | $< 0.96$ | 1.03(139) | $1.60^{+0.07}_{-0.07}$ | 0.224 | 1.02(140) |
| GIS2 | $1.57^{+0.40}_{-0.29}$ | $< 4.61$ | 0.98(62) | $1.49^{+0.16}_{-0.15}$ | 0.224 | 0.96(63) |
| GIS+SIS | $1.63^{+0.14}_{-0.12}$ | $< 0.66$ | 1.19(204) | $1.64^{+0.07}_{-0.06}$ | 0.224 | 1.19(205) |
| *PKS 2149-306* | | | | | | |
| SIS0+1 | $1.55^{+0.07}_{-0.06}$ | $0.73^{+0.23}_{-0.23}$ | 1.01(343) | $1.44^{+0.02}_{-0.03}$ | 0.245 | 1.06(344) |
| GIS2+3 | $1.54^{+0.09}_{-0.09}$ | $< 1.25$ | 1.04(338) | $1.50^{+0.04}_{-0.03}$ | 0.245 | 1.04(339) |
| GIS+SIS | $1.57^{+0.05}_{-0.05}$ | $0.59^{+0.20}_{-0.19}$ | 1.19(684) | $1.48^{+0.04}_{-0.02}$ | 0.245 | 1.21(685) |
| | Thermal Bremsstrahlung | | | | | |
| | kT[keV] | $N_H^a$ | $\chi^2_{red}$ (dof) | kT[keV] | $N_{H,gal}$ | $\chi^2_{red}$ (dof) |
| *PKS 0332-403* | $10.4^{+4.3}_{-2.6}$ | $< 0.48$ | 1.23(134) | $10.0^{+2.6}_{-2.0}$ | 0.145 | 1.23(135) |
| *PKS 0537-286* | $33.5^{+12.0}_{-7.6}$ | $< 0.18$ | 1.22(204) | $29.7^{+7.1}_{-5.1}$ | 0.224 | 1.25(205) |
| *PKS 2149-306* | $45.8^{+9.8}_{-6.6}$ | $< 0.24$ | 1.23(684) | $41.4^{+4.2}_{-3.6}$ | 0.245 | 1.24(685) |

[a] In units of $10^{21}$ cm$^{-2}$.
All errors and upper limits are 90% confidence values for 2 interesting parameters.

More complicated continuum emission models are not required. Nevertheless, we determined the upper limits for the Fe-K emission line for all sources. The 90% upper limits (for 1 interesting parameter, $\Delta\chi^2 = 2.706$) on the equivalent width of a narrow ($\sigma = 0$) Fe-K emission line at 6.4 keV in the rest frame of the sources are 298 eV, 195 eV, and 19 eV for PKS 0332-403, PKS 0537-286, and PKS 2149-306, respectively.

We also tried to model the ASCA spectrum of the strongest source in our sample, PKS 2149-306, with a reflection model in order to investigate if the flatness of the rest frame $\sim$1.5-33 keV spectrum of this quasar could possibly be caused by the presence of a reflection component. However, the reflection model predicts a Fe-K 6.4 keV emission line with an EW of $\sim$ 160 eV, much higher than the 90% upper limit of 19 eV which we determined above. We therefore conclude that a reflection model is inconsistent with the X-ray spectrum of PKS 2149-306.

## 4. Comparison with ROSAT

All four objects of the sample were detected in the ROSAT All-Sky Survey. In addition, PKS 0537-286 was observed in a pointed ROSAT PSPC observation on Sept. 28/29 1992 with a total exposure of $\sim$ 9400 seconds.

In Table 4 we give the unabsorbed ASCA fluxes and luminosities for the objects in our sample and compare them to the ROSAT values. The luminosity in the 2–10 keV rest frame energy band is listed in column 2, whereas columns 3 and 4 give the unabsorbed ASCA fluxes in the 0.8–8 keV and the 0.1–2.4 keV energy range, respectively. The ASCA 0.1–2.4 keV flux is calculated by extrapolating the best fit model to 0.1 keV. Finally, we list in column 5 the corresponding ROSAT PSPC fluxes as obtained from the ROSAT All-Sky Survey and from the pointed observation of PKS 0537-286. All fluxes and luminosities are calculated assuming the best fit ASCA photon indices and Galactic absorption. For PKS 1614+051, we assumed a photon index of $\Gamma = 1.6$.

The errors of the ROSAT All-Sky Survey fluxes are $\sim$ 25% from photon statistics alone. Thus, within the errors,

**Table 4.** X-ray fluxes and luminosities

| Object | $L_{2-10keV}$ erg s$^{-1}$ | $f_x$ † 0.8-8 | $f_x$ 0.1-2.4 | $f_x$ PSPC |
|---|---|---|---|---|
| PKS 0332−403 | $1.6 \cdot 10^{46}$ | 0.89 | 1.08 | 1.38 |
| PKS 0537−286 | $2.0 \cdot 10^{47}$ | 1.52 | 1.16 | 1.04 |
| PKS 1614+051 | $8.2 \cdot 10^{46}$ | 0.20 | 0.32 | 0.49 |
| PKS 2149−306 | $5.7 \cdot 10^{47}$ | 9.87 | 6.97 | 7.78 |

† All fluxes are in units of $10^{-12}$ erg cm$^{-2}$ s$^{-1}$.

the ROSAT fluxes and the ASCA fluxes, extrapolated to the ROSAT energy band, are in reasonable agreement.

The ROSAT spectrum of PKS 0537-286 is consistent with the ASCA results. The best fit power-law index is $\Gamma = 1.48 \pm 0.42$ (90% confidence for two interesting parameters). There is no evidence for excess absorption in the PSPC spectrum of PKS 0537-286. The best fit $N_H$ value ($0.32^{+0.18}_{-0.14} \times 10^{21}$ cm$^{-2}$) is not significantly higher than the Galactic $N_H$ ($0.224 \times 10^{21}$ cm$^{-2}$).

Only ∼22 and ∼10 photons were accumulated in the ROSAT All-Sky Survey for PKS 0332-403 and PKS 1614+051, respectively. Therefore a detailed spectral analysis is impossible.

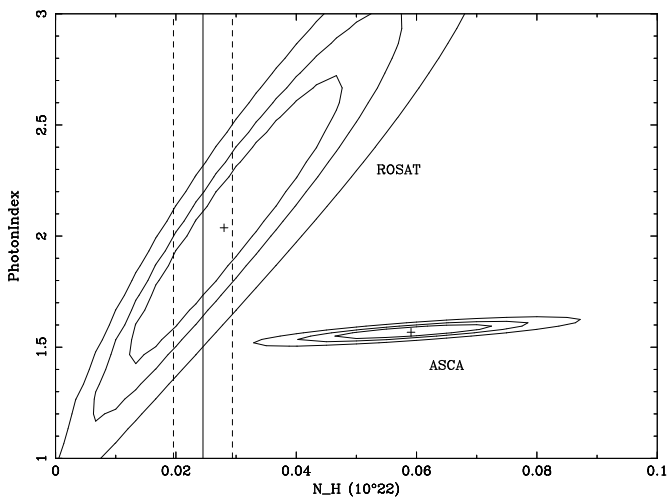

**Fig. 2.** Contour plot of photon index $\Gamma$ versus fitted $N_H$-value for PKS 2149-306. The contours correspond to 68%, 90% and 99% confidence, respectively. Also shown as a vertical line is the value of Galactic absorption according to Stark et al. (1992). The dashed lines represent a 20% variation of the $N_H$ value.

Fortunately, PKS 2149-306 yielded a sufficient number of photons in the ROSAT All-Sky Survey to allow a rough determination of the spectral parameters in the 0.1-2.4 keV energy band. In total, ∼165 photons were ob-ple power-law fit plus absorption are $\Gamma = 2.04^{+1.01}_{-0.87}$ and $N_H = 0.28^{+0.30}_{-0.22} \times 10^{21}$ cm$^{-2}$ ($\chi^2_{red} = 0.83$). Even in view of the small number of photons in the ROSAT All-Sky Survey observation of this source, we are led to conclude that there is no evidence for absorption in excess of the Galactic value of $N_H = 0.245 \times 10^{21}$ cm$^{-2}$ in the PSPC spectrum. This is in apparent contradiction to the ASCA results. Assuming the best fit ASCA spectral parameters ($\Gamma = 1.57$, $N_H = 0.59 \times 10^{21}$ cm$^{-2}$), gives an unacceptable fit with $\chi^2_{red} = 3.37$. However, fixing only the *photon index* to the best fit ASCA value, results in an acceptable fit ($\chi^2_{red} = 0.86$) with $N_H = (0.16 \pm 0.06) \times 10^{21}$ cm$^{-2}$, which is roughly consistent with Galactic absorption.

In Figure 2 we show the 68%, 90% and 99% confidence contours of the spectral parameters for ASCA and ROSAT. Even the 99% contours do not overlap. Further, it can be seen from the shape of the contours that the two spectra are complementary in the sense that the ASCA data are more sensitive to the spectral index, while the ROSAT data are more sensitive to absorption.

An explanation for these contradictory results could be provided by variable absorption. In this case the absorption must have changed by at least a factor of two ($\Delta N_H \sim 4 \times 10^{20}$ cm$^{-2}$) between the two observations, i.e. within ∼ 30 months (∼ 9 months rest frame). The comparably short timescale then indicates that the excess absorption is most likely intrinsic. In this case, the change in absorption corresponds to $\Delta N_H \sim 5.5 \times 10^{21}$ cm$^{-2}$.

Variable soft emission is probably inappropriate to explain the results, because it is difficult to mimic excess absorption in the ASCA spectrum by this effect. If there were two emission components, a variable soft and a second harder one extending to higher energies, they have to be distinct in the sense that there is no contribution from the hard component below ∼ 0.5 keV (∼ 1.7 keV rest frame). Though not impossible, this seems rather unlikely.

The ASCA result could also be affected by the reduced capability of the detectors to determine accurately low values of $N_H$ ($< 10^{21}$ cm$^{-2}$), and detector calibration uncertainties at low energies might play a role as well. The reason why we don't see this instrumental effect in PKS 0537-286, which exhibits an otherwise similar spectral behaviour, is probably the much lower intensity of this source below ∼ 1 keV compared to PKS 2149-306.

## 5. Summary and discussion

We presented ASCA observations of four high redshift radio-loud quasars with 1.445<z<3.21. Three sources (PKS 0332-403, PKS 0537-286, PKS 2149-306) were sufficiently bright to allow a detailed spectral analysis. The fourth and most distant quasar, PKS 1614+051, was detected as well, but unfortunately the number of source ters.

Within the errors the derived spectral indices are compatible with previous ASCA results for high redshift radio-loud quasars. Serlemitsos et al. (1994) report $\Gamma = 1.63^{+0.08}_{-0.09}$ and $\Gamma = 1.70^{+0.20}_{-0.18}$ for the radio-loud quasars PKS 2126-158 and PKS 0438-436, respectively. Elvis et al. (1994b) give $\Gamma = 1.63 \pm 0.03$ as best fit result to the GIS spectrum of S5 0014+813. Further, the X-ray spectra of high redshift radio-loud quasars remain flat at lower energies, as can be inferred, for example, from our ROSAT PSPC data of PKS 0537-286 and PKS 2149-306 and from Bechtold et al. (1994b), who derive a mean photon index of $1.59 \pm 0.06$ for five radio-loud high redshift quasars from the analysis of PSPC hardness ratios. PKS 0332-403 might exhibit a slightly steeper X-ray continuum than the other high redshift quasars. We note however, that PKS 0332-403 is also the closest object in our sample. Therefore it is conceivable that we are seeing the intermediate part of the spectrum between the steep low energy continuum (e.g. Wilkes & Elvis 1987, Brinkmann et al. 1995) and the flat spectrum at higher energies.

ASCA and ROSAT spectra of high redshift quasars can be compared to the 2-20 keV *Ginga* spectra of low redshift quasars, as the intrinsically observed energy range is approximately the same in both cases. The 90% confidence range of the spectral index for radio-loud quasars in Williams et al. (1992) is $1.58 < \Gamma < 1.86$, thus in good agreement with the results presented in this paper and the ASCA and ROSAT results listed above. This implies that the spectra of radio-loud quasars have not changed significantly since z∼3, although the average luminosities in the 2–10 keV energy band for the low redshift quasars of Williams et al. (1992) and the high redshift quasars dicussed in this paper differ by a factor of ∼50.

In the presently discussed evolutionary scenarios, the mass of the central black hole is either believed to increase with time by accretion (if quasars are a long-lived phenomenon) or to decrease if quasars are short-lived and if they form in processes as proposed for example by Haehnelt & Rees (1993). Thus, the absence of evolution in the X-ray spectra of radio-loud quasars at least indicates that the shape of the X-ray continuum is not directly related to the mass of the central black hole.

In the ASCA spectra of PKS 0332-403 and PKS 0537-286 we find no evidence for absorption in excess of the Galactic value. In the case of PKS 2149-306 we report excess absorption above the Galactic $N_H$ value with more than 99% confidence. However, the low energy ROSAT observation of PKS 2149-306 is inconsistent with the amount of absorption derived from the ASCA data for this source. These apparently contradictory findings can either be explained by variable absorption, most likely intrinsic to the source ($\Delta N_H \sim 5 \times 10^{21}$ cm$^{-2}$ within 9 months in the restframe of the quasar), or by a variable soft X-ray component. The latter possibility seems unlikely, however, since hard component below 0.5 keV in order to mimic excess absorption in the ASCA spectrum. Further, currently existing calibration uncertainties of the ASCA detectors at low energies might play a role as well.

Elvis et al. (1994a) find excess absorption in the ROSAT PSPC spectra of 2 (out of 3) radio-loud quasars (PKS 0438-436 and PKS 2126-158) and claim absorption to be a common feature in high redshift radio-loud quasars. In a comparative study of the PSPC spectra of radio-loud and radio-quiet high redshift quasars, Bechtold et al. (1994b) claim to find additional absorption in 3 (out of 5) sources, including the two objects of Elvis et al. (1994a). Together with the four sources discussed in this paper, the X-ray data of roughly a dozen of high z radio-loud quasars have been published. The three objects of Elvis et al. (1994a) and Bechtold et al. (1994b) are up to now the only sources where excess absorption has been unambigiously detected. This indicates that excess absorption is probably not common in high redshift radio-loud quasars.

In Figure 3 we present the spectral energy distributions (SED) of the four quasars in their respective rest frames and compare them to the average SED of radio-loud quasars determined by Elvis at el. (1994c). The average SED has been normalized to the ROSAT X-ray flux at 1 keV. The optical magnitude has been corrected for Galactic reddening using the $N_H$ values and assuming standard gas-to-dust ratios.

As already noted by Bechtold et al. (1994a), the average SED for radio-loud quasars from Elvis et al. (1994c) tends to underestimate the radio fluxes. This is due to the fact that the Elvis et al. (1994c) sample consisted of quasars bright enough in X-rays to do meaningful spectral fits with the *Einstein* IPC data. The effect may not be very strong in our sample, however, as the selected objects are also X-ray bright.

Keeping this in mind, PKS 2149-306 and PKS 0537-286 still exhibit unusual spectral energy distributions in the sense that their X-ray and radio fluxes are in reasonable agreement with the medium SED, whereas the optical fluxes are almost one order of magnitude lower. The error bars on the fluxes are comparably small and cannot explain the deviations. As the data points are not taken simultaneously, variability might play a role, but most likely cannot account for the whole effect either. Alternatively, the X-ray and radio fluxes could be unusually high compared to the optical flux, for example caused by beamed radio emission and hence increased X-ray flux via inverse–Compton emission.

Finally, we note the high radio flux of PKS 0332-403 relative to the X-ray emission. The alternative explanantion, that the X-ray flux could be in low state compared to the radio band seems unlikely, since also the optical flux fits well to the average SED. In view of the fact that PKS 0332-403 is known to be highly polarized in the optical (Impey & Tapia 1990), an interpretation in terms of enhanced radio emission due to relativistic beaming in the radio jet seems plausible. However, current models for the X-ray emission of quasars (e.g. Wilkes & Elvis 1987) associate a flat X-ray component with the radio emission (i.e., synchrotron self-Compton) in order to explain the flatter X-ray spectra in radio-loud quasars compared to radio quiet QSO. Thus an increased radio flux should lead to a flattening of the spectrum, contrary to the steep X-ray spectrum observed in PKS 0332-403.

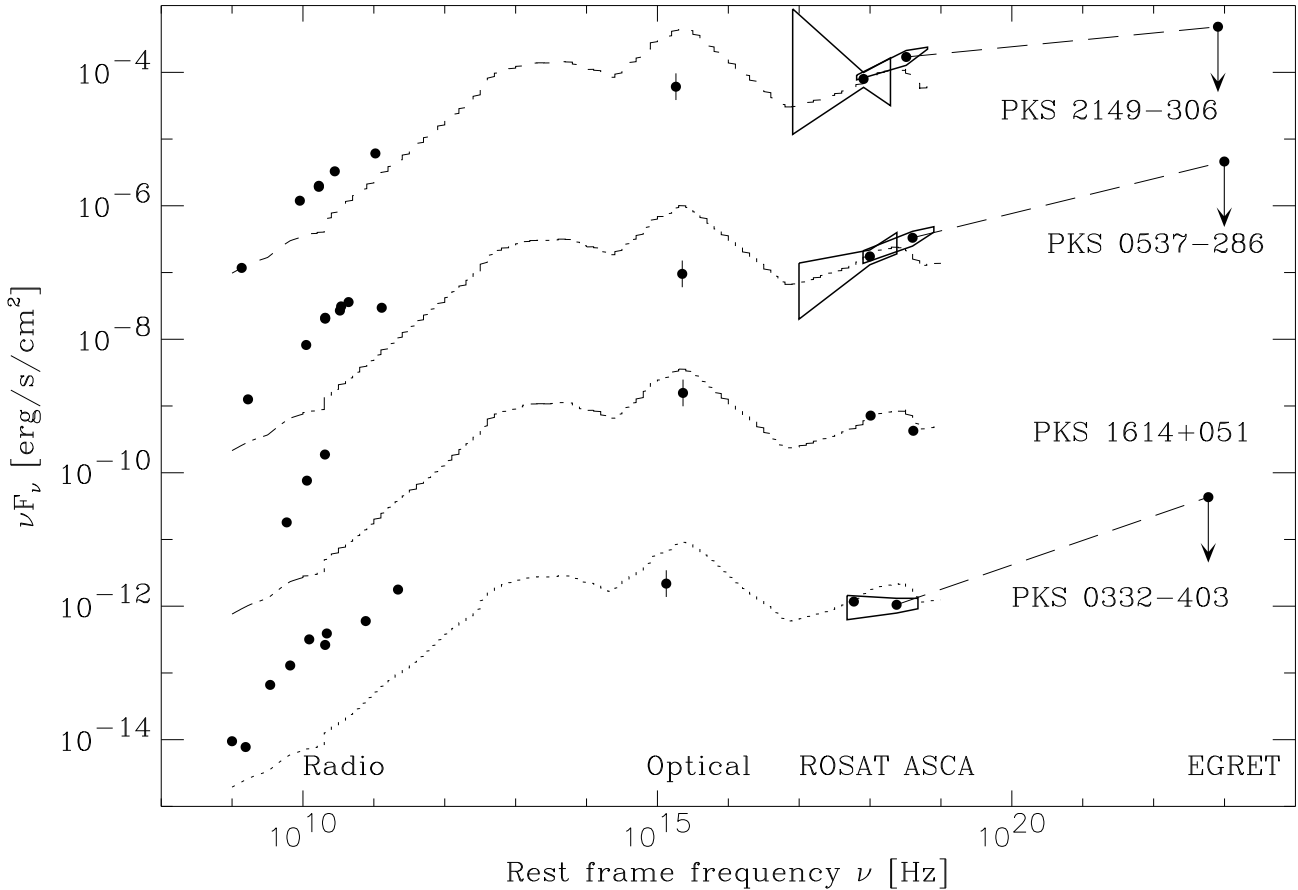

**Fig. 3.** Spectral energy distributions (SED). The objects are ordered according to the 2-10 keV luminosity with the brightest source at the top. For clarity, the fluxes are shifted upwards by two, four and six orders of magnitude, respectively. The SED's are compared to the average SED for radio-loud quasars (Elvis et al. (1994c) normalized to the ROSAT X-ray flux. The error bars on the optical data points correspond to $\Delta m = \pm 0.5 mag$.

*Acknowledgements.* JS and WB acknowledge financial support from the RIKEN–MPG exchange program. They also would like to thank their collegues of the Cosmic Radiation Laboratory at RIKEN for hospitality and support during their stay at the Institute. This research has made use of the NASA/IPAC Extragalactic Data Base (NED) which is operated by the Jet Propulsion Laboratory, California Institute of Technology, under contract with the National Aeronautics and Space Administration.

## References

Aldcroft T., Elvis M., McDowell J., Fiore F., 1994, ApJ 437, 584

Bechtold J., Elvis M., Fiore F., et al., 1994a, AJ 108, 374

Bechtold J., Elvis M., Fiore F., et al., 1994b, AJ 108, 759

Boyle B.J., Griffiths R.E., Shanks T., Stewart G.C., Georgantopoulos I., 1993, MNRAS 260, 49

Brinkmann W., Siebert J., Reich W., et al., 1995, A&AS 109, 147

Burke B.E., Mountain R.W., Harrison D.C., et al., 1991, IEEE Trans. ED-38, 1069

Canizares C.R., White J.L., 1989, ApJ 339, 27

Della Ceca R., Maccacaro T., 1991, in Crampton D., ed., The Space Distribution of Quasars, ASP Conference Series 21, p.150

Elvis M., Fiore F., Wilkes B., McDowell J., 1994a, ApJ 422, 60

Elvis M., Matsuoka M., Siemiginowska A., et al., 1994b, ApJ 436, L55

Elvis M., Wilkes B.J., McDowell J.C., et al., 1994, ApJS 95, 1

551

Fukazawa Y., Makishima K., Ebisawa K., et al., 1994, PASJ 46, L141

Haehnelt M., Rees M., 1993, MNRAS 263, 168

Impey C.D., Tapia S., 1990, ApJ 354, 124

Lightman A.P., White T.R., 1988, ApJ 331, 57

Morrison R., McCammon D., 1983, ApJ 270, 119

Ohashi T., Makishima K., Ishida M., et al., 1991, Proc. SPIE 1549, 9

Serlemitsos P., Yaqoob T., Ricker G., et al., 1994, PASJ 46, L43

Stark A.A., Gammie C.F., Wilson R.W., et al., 1992, ApJS 79, 77

Tanaka Y., Inoue H., Holt S.S., 1994, PASJ 46, L37

Trümper J., 1984, Physica Scripta 7, 209

Wilkes B.J., Elvis M., 1987, ApJ 323, 243

Williams O.R., Turner M.J.L., Stewart G.C., et al., 1992, ApJ 389, 157